# Coupling Coordinated Development among Digital Economy, Regional Innovation and Talent Employment
## 一A case study of Hangzhou Metropolitan Circle, China

Qiu Luyi*
*Corresponding author, Business School, Universiti Kuala Lumpur, 1016 Jalan Sultan Ismail, 50250 Kuala Lumpur, Malaysia.
Department of International Trade and Economics, Ningbo Polytechnic, Ningbo 315800, China.
Email: 0000036@nbpt.edu.cn
Sharina Osman
Business School, Universiti Kuala Lumpur，1016, Jalan Sultan Ismail, 50250 Kuala Lumpur, Malaysia.
Email: sharina@unikl.edu.my.

**Abstract:** Coordination development across various subsystems, particularly economic, social, cultural, and human resources subsystems, is a key aspect of urban sustainability that has a direct impact on the quality of urbanization. Hangzhou Metropolitan Circle composing Hangzhou, Huzhou, Jiaxing, Shaoxing, was the first metropolitan circle approved by National Development and Reform Commission (NDRC) as a demonstration of economic transformation. To evaluate the coupling degree of the four cities and to analyze the coordinative development in the three systems (Digital Economy System, Regional Innovation System, and Talent Employment System), panel data of these four cities during the period 2015-2022 were collected. The development level of these three systems were evaluated using standard deviation and comprehensive development index evaluation. The results are as follows: (1) the coupling coordination degree of the four cities in Hangzhou Metropolitan Circle has significant regional differences, with Hangzhou being a leader while Huzhou, Jiaxing, Shaoxing have shown steady but slow progress in the coupling development of the three systems; and (2) the development of digital economy and talent employment are the breakthrough points for construction in Huzhou, Jiaxing, Shaoxing. Related suggestions are made based on the coupling coordination results of the Hangzhou Metropolitan Circle.

**Key words:** Digital economy.   Regional innovation. Talent Employment.   Coupling coordination degree.   Hangzhou Metropolitan Circle

# 1. Introduction

As a type of regional spatial structure distinct from others, a metropolitan circle is the result of urban development up to a particular point (Galster& Killen, 2007). It is made up of one or more core cities, as well as connected towns and circles with strong social and economic linkages to the core and integration inclinations (Fang&Yu,2017). China's urbanization has continued to improve since reform and opening up, and the economic, social, and cultural integration process among cities in economic developed regions such as the Yangtze River Delta, the Pearl River Delta, and the Beijing-Tianjin-Hebei region has accelerated, resulting in the formation of a number of metropolitan circles (Liu & Zhang ,2021). Many cities have begun to actively build metropolitan circles with large cities as the core and surrounding small and medium-sized cities as members in order to strengthen regional economic collaboration, achieve organic resource integration, and improve overall city competitiveness (Sahana & Hong ,2018).

Hangzhou Metropolitan Circle is a key portion of the Yangtze River Delta urban agglomeration and is located in the Yangtze River Delta economic zone's south. Hangzhou Metropolitan Circle has a spatial pattern with Hangzhou city as the core, Huzhou, Jiaxing, Shaoxing as the sub center, 20 neighboring counties including Deqing, Anji, Haining, Tongxiang, Chunan, Jiande, Changxing as the close layer, and a large number of small towns as the surrounding layer. The structure of the Hangzhou Metropolitan Circle is relatively clear, and it has a reasonable arrangement of large cities, medium-sized counties, and small towns, making it one of China's regions with the highest degree of industrial agglomeration and coordination (Shen,2019). Since 2015, the Hangzhou Metropolitan Circle has promoted industrial space optimization, industrial transformation and upgrading, and regional economic innovation and development by constructing a number of feature towns with industrial and humanities features, eco-tourism, and community functions (Zhu & Chen, 2022). During the 14th Five-Year Plan period, the Hangzhou Metropolitan Circle capitalized on digital economy to achieve integrated digital and urban economic development.

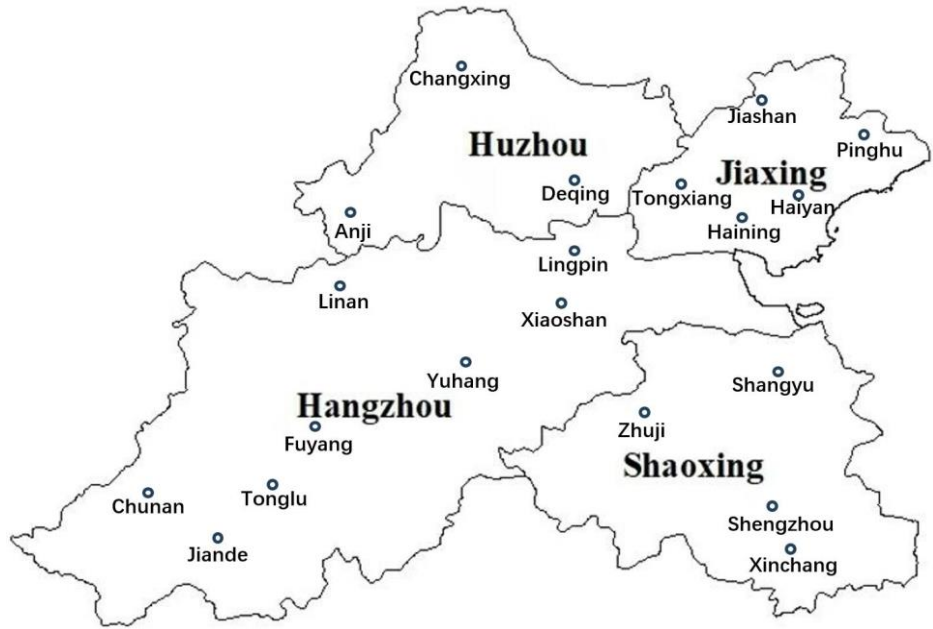

Figure 1. The structure of Hangzhou Metropolitan Circle

Under the spatial pattern of "big city and small town", the Hangzhou Metropolitan Circle network presents a three circles layer of industrial layout. Firstly, the core circle focuses on information technology and software, e-commerce, finance, cultural and creative industries, and so on. Secondly, the inner circle prioritizes intelligent manufacturing, pharmaceutical and chemical industry, textile and clothing, and the rubber and plastic industry, and so on. Thirdly, the outer circle develops chemical smelting, leather, agriculture, rural tourism and other related industries according to the different advantages and characteristics of each county and small towns (Figure 2). Under the spatial pattern of "big cities and small towns," a mechanism to strengthen interactions and cooperation among metropolitan circles is critical (Shen,2019). As a result, studying the coordinated development of industry, technology, and human resources in the Hangzhou Metropolitan Circle will help to establish a coordination guarantee mechanism for the metropolitan circle's development (Duarte et al., 2021). Finally, we hope to achieve the optimal allocation and sustainable development of industry, technology, and human resources in the Hangzhou Metropolitan Circle.

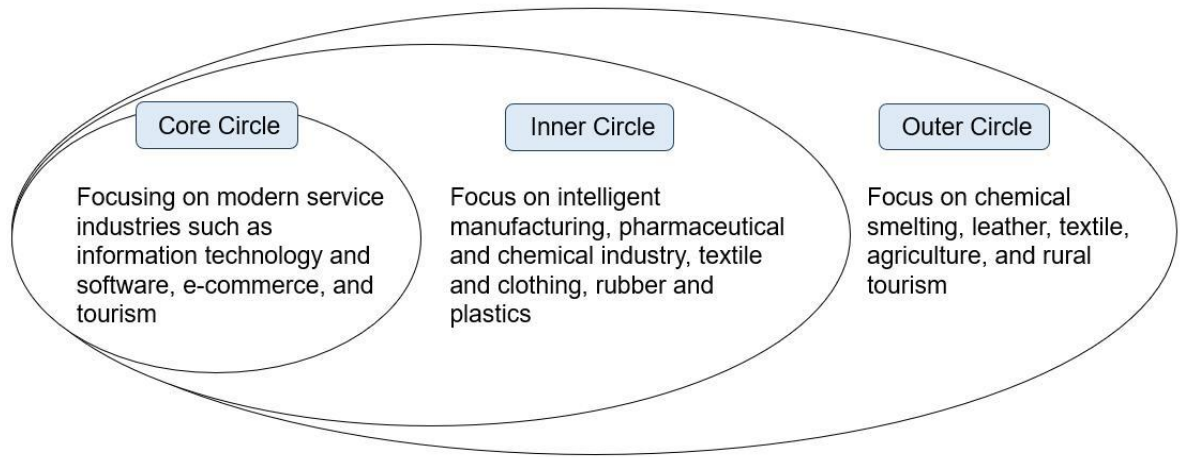

Figure 2. The spatial distribution of industrial circles in the Hangzhou Metropolitan Circle

## 2. Literature reviews

At the moment, domestic and international researchers' research on metropolitan circle focuses mostly on concepts, scope, population size, urban relationships, innovative development, and so on. There is not much literature on the coordinated development of metropolitan circle Hou et al. (2020) investigated changes in the coordinated development of urbanization in the Chengdu metropolitan circle, as well as the coordination of urbanization with economic and environmental growth. They argued that the Chengdu metropolitan circle's urbanization process should pay greater attention to the population-economy-environment coordinated development model. In the study of the Hangzhou Metropolitan Circle, Zhu (2017) showed that the digital economy, as the core industry of the Hangzhou Metropolitan Circle, mainly reflects the agglomeration of digital economic elements from a time and space perspective. Shen et al. (2015) believed the digital economy space is a gathering region with significant resource integration and radiation functions produced by cooperation, interaction, and innovation. However, Valenduc & Vendramin (2016); Pan et al. (2022) showed that digital economy elements can be divided into hardware and software, of which "soft elements" such as digital technology talents, digital industry policy and digital technology service are more critical, and can provide policy, innovation, and service support for the agglomeration development of the digital economy space. Banalieva & Dhanaraj (2019); Li et al. (2022) considered that spatial agglomeration theory of urban economics emphasizes the influence of Technology Spillover on regional innovation, he believed that knowledge and technology spillovers are conducive to industrial agglomeration and expanding industrial agglomeration can strengthen the spillover of knowledge and technology. Kogan et al. (2017) believed based on the fact that technological innovation has begun to show a greater spatial agglomeration effect, the spatial spillover of knowledge and technology has become an important factor affecting regional innovation with the rapid development of network, communication, and information technology. Alrajhi & Aydin (2019) showed universities, research institutes, and technological R&D centers contribute significantly to the intellectual development of this digital economic space. According to the research of Guo &Tao (2022) , new industries, new formats and new models of small towns are constantly emerging through technological innovation, efficient networks, and convenient transportation, which puts forward higher requirements for the human capital of small towns. Kong & Chen (2019) argued that it is necessary to give full play to the characteristic advantages of "small but refined" local service industries, highlight the specialized scale effect and diversified integration interaction, broaden the talent employment channels of service industry, improve the capacity of employment absorption, and narrow the gap with big cities.

From the perspective of quantitative analysis, Billings & Johnson (2012) have utilized the Location Quotient (LQ) to calculate the industrial agglomeration index of different spaces to measure the degree of industrial agglomeration. For example, the spatial agglomeration degree of an industry can be evaluated through the designated relationship between the employment population of specific industries and the total population of the region. Although this method of measurement provides a useful perspective for the study of industrial agglomeration, LQ focuses on reflecting the degree of specialization of an industrial sector, and the index it measures is not comprehensive enough, lacking in multi-dimensional, multi-level and multi form analysis.

According to previous research, digital economy, regional innovation and talent employment in metropolitan circles are three keys to promoting the upgrading and high-quality development of a regional economy. Therefore, this paper attempts to analyze the coupling and coordinated development of Hangzhou Metropolitan Circle from the perspective of coupling and coordination with an aim to better plan the overall development of Hangzhou Metropolitan

Circle.

## 2. METHODOLOGY

### 2.1 Construction of a Coupling Coordination Model of Digital Economy, Regional Innovation and Talent Employment

Digital economy, regional innovation, and talent employment all play essential roles in supporting and encouraging regional economic growth in the process of forming and developing industrial space in metropolitan circles (Dong et al., 2020). In the process of implementing the regional innovation and development strategy, Zhu et al. (2020) considered the Hangzhou Metropolitan Circle "promotes city by industry, integrates industry with city," builds a pattern of coordinated development of large, medium, small cities, and small towns, and forms various types of digital economy carriers, such as development zones, industrial parks, entrepreneurial parks, and feature towns. Under the guidance of the digital economy, a large number of small towns have accelerated their integration into the metropolitan network and participate in the division of labor and cooperation based on their own competitive advantage, and the three-dimensional spatial pattern of the "Hangzhou Metropolitan Circle" is gradually emerging (see Figure 3).

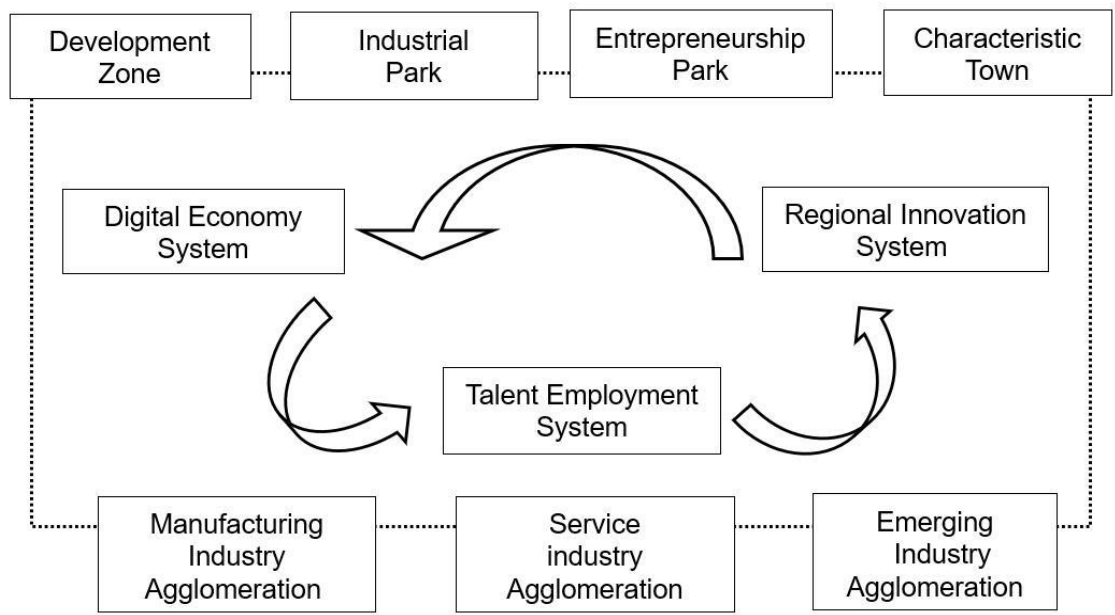

Figure 3. Correlation between digital economy, regional innovation and talent employment in Hangzhou Metropolitan Circle

Due to the complex coordination relationship of multi-dimensional, multi-level and multi-form among the three systems of digital economy, regional innovation and talent employment, and to fully reveal the interaction between the three systems, this study selects four cities in the Hangzhou Metropolitan Circle before expansion, Hangzhou, Shaoxing, Jiaxing and Huzhou, as objects to explore the coupling and coordinated development of digital economy , regional innovation and talent employment in the Hangzhou Metropolitan Circle under the spatial pattern of big cities and small towns. A comprehensive evaluating index system was established with 6 subjects and 24 indicators for evaluating the three systems, i.e., digital economy, reginal innovation and talent employment. The Mean Squared Deviation (MSD) method was used to determine the weight of each indicator and subsystem. The Comprehensive Development Index was used to measure the development of the three systems based on weight determined by MSD. Coupling Coordination Degree Model (CCDM) was used to measure the coupling coordination degree among the three systems. By doing so, policy implications for the coordinated development of Hangzhou Metropolitan Circle can be offered.

## 2.2 Index System Establishment

Zhu (2017) believed that digital industry agglomeration is affected by many factors, and that we usually select indicators related to digital economy infrastructure, including the number of mobile phone users, the number of internet connections, telecommunications industry income, are frequently used as metrics to assess the degree of agglomeration of the digital economy. In this paper, we also select the regional indicators related to the scale of digital industry agglomeration, including the number of foreign investment projects in the digital economy, cultural creativity, service outsourcing and other tertiary industries, as well as the freight volume and express delivery quantity reflecting the scale of regional digital economy development at the logistics data level.

In recent years, the Hangzhou Metropolitan Circle has strengthened the application of new knowledge and new technology in the process of transformation and upgrading of small and medium-sized enterprises, and the human, material and financial resources invested in scientific research and experimental development activities have increased significantly (Zhu et al., 2020; Chen,2015). Therefore, in terms of regional innovation input, this paper selects four indicators related to R & D development to measure the scale and intensity of scientific and technological innovation activities in a region. In terms of innovation output, Pang et al. (2019) and other scholars listed the number of science and technology business incubators as an important evaluation index. In addition, considering the time lag and other reasons, this paper selects the number of patent applications and the output value rate of new products as other indicators of innovation output.

To evaluate the relationship between talent employment and regional economic development scientifically and effectively, we take the number of graduates from colleges and universities, vocational schools, and technical schools as the main indicators to measure the education scale and regional talent employment level of a region (Fernandes & O'Sullivan,2022). The reason why the number of graduates from secondary vocational schools and technical schools is also included in the evaluation system is that such schools play a very important role in exporting "blue-collar" employment groups such as junior and intermediate technical personnel and skilled workers to the society with the development of vocational education (Liu & Li,2019). In addition, the world's largest professional social networking platform (LinkedIn) shows that the most popular employment fields of college graduates in Hangzhou Metropolitan Circle are mainly concentrated in information technology, computer services, software development, business services, finance, and other tertiary industries. Therefore, in the talent employment system, we take the number of employees in these popular employment fields as the main indicator. Combining these with the development basis and scale of the digital economy in the Hangzhou Metropolitan Circle, the input and output of regional innovation, and the scale and quantity of talent employment, a comprehensive evaluation system which includes three levels and 24 indicators of digital economy, regional innovation and talent employment is constructed (**Table 1**).

The sources of research data must follow the principles of reliability, quantification, comparability, and stability. The data in this paper are taken mainly from the statistical yearbooks and bulletins of four cities in 2015-2022.The data needs to be standardized before weight determination and data analysis as the dimensions and units of each index are different. The formula for standardization is as follows. Formula (1) is for index with positive effect and formula (2) is for the index with negative effect.

$$X'_{ij} = \frac{X_{ij} - X_{i\min}}{X_{i\max} - X_{i\min}} \quad (1) \qquad X'_{ij} = \frac{X_{\max} - X_{ij}}{X_{i\max} - X_{i\min}} \quad (2)$$

In the above formula, $X'_{ij}$ is standardized values. $X_{ij}$ is the original i value of indicator j in the year of I; and $X_{i\max}$ $X_{i\min}$ represent the maximum and minimum values of the original data in the evaluation period, respectively.

Table 1. Index Weight of Comprehensive Evaluation System of Digital Economy, Regional Innovation and Talent Employment

| Systems | Subsystems | Indicators | Effect | Weight |
|---|---|---|---|---|
| **Digital Economy Systems** | Foundation of digital industry | Revenue of Telecommunication industry | + | 0.159 |
| | | Number of mobile phone users at the end of the year | + | 0.167 |
| | | Number of broadband services users at the end of the year | + | 0.121 |
| | | Foreign direct investment projects in the tertiary industry | - | 0.123 |
| | Scale of digital economy | Scale of Cultural and creative industries | + | 0.109 |
| | | Scale of Information economy industry | + | 0.116 |
| | | Freight transport volume | + | 0.125 |
| | | Express delivery volume | + | 0.080 |
| **Regional Innovation Systems** | Innovation investment | R & D investment of the whole society | + | 0.135 |
| | | Personnel Research and experimental development activities | + | 0.122 |
| | | Ratio of R & D expenditure to GDP | + | 0.182 |
| | | Number of scientific and technological institutions in Enterprises | + | 0.122 |
| | Innovation output | Number of patents granted | + | 0.142 |
| | | Number of research and experimental development projects of industrial enterprises above designated size | + | 0.091 |
| | | Output value rate of new products | + | 0.114 |
| | | Number of technology business incubators | + | 0.091 |
| **Talent Employment Systems** | Scale of education | Number of students in Colleges | + | 0.120 |
| | | Number of college graduates | + | 0.187 |
| | | Number of graduates from vocational schools and technical schools | + | 0.082 |
| | Talent employment | Proportion of employment in tertiary industry | + | 0.092 |
| | | Employment in information transmission, computer services and software | + | 0.093 |
| | | Employment in industry of leasing and business services | + | 0.085 |
| | | Employment in Financial Industry | + | 0.132 |
| | | Employment in scientific research and technical services | + | 0.129 |

## 2.3 Index Weight Calculation

To avoid the subjectivity of determining weights, this paper uses the Mean Squared Deviation method based on information entropy to determine index weights. Entropy is a scientific measure of the degree of system chaos by the variability of indicators. The standard deviation and variability of indicators are consistent. By calculating the standard deviation of each index, the larger the standard deviation of the same index, the greater the variation of the index, the larger the information provided, and the greater the role in the comprehensive evaluation, and vice versa (Wang, 1999). The calculation process of mean square deviation and weight is as follows:

$$\partial_j = \sqrt{\frac{1}{n}\sum_{i=1}^{n}\left(x_{ij}' - \overline{x_{ij}}\right)^2}, \overline{x_{ij}} = \frac{1}{n}\sum_{i=1}^{n} x_{ij}'$$

$$W_j = \partial_j / \sum_{j=1}^{m} \partial_j$$

## 2.4 The Coupling Coordination Degree Model

As regional digital economy, regional innovation and talent employment are examples of complex systems engineering and the relationship between each subsystem is complex, it is difficult to analyze its operating mechanism based on experience. Therefore, this paper uses a Coupling Coordination Degree Model (CCDM) to analyze the correlation between the three and how the three systems interact to influence each other. The coupling coordination model originated from physics, developed to study the interaction between two or more systems and was then gradually applied to many fields such as geography, economy, science and technology, tourism and so on. By referring to the model of capacity coupling coefficient in physics and coupling three different but interactive systems, regional digital economy, regional scientific and technological innovation, and regional talent employment, which can reflect benign interaction and interdependence among subsystems, the CCDM of the three systems can be built and expressed as follows:

$$C = \left\{ \frac{f(E) \times g(U) \times m(X)}{(f(E) + g(U) + m(X))^3} \right\}^{1/3}$$

In this formula, C represents the coupling degree among digital economy, regional innovation, talent employment system. F (E) is the comprehensive development Index (CDI) of regional digital economy system; g (U) is the CDI of regional science and technology innovation system; M (x) is the CDI of regional talent employment system. The value range of C is 0-1, that is to say, when C = 0, the three systems are in the worst correlated and completely disordered state; when C = 1, the three systems are in the most correlated and orderly state. To more accurately judge the degree of harmony among the three systems of digital economy, regional innovation and talent employment, and reveal the dynamic equilibrium development state of those three systems, the coupling coordination model can be further constructed as follows:

$$T = \alpha f(E) + \beta g(U) + \chi m(X)$$
$$D = \sqrt{C \times T}$$

In the formula, T represents the comprehensive evaluation index of the overall synergistic effects of system f (E), system g (U), and system m (X); D represents the degree of coupling coordination of system f(E), system g(U), and system m(X), and the higher the D value, the better the coordination among the three systems, and vice versa, the worse the correlation and coordination among the three systems. Considering the equal importance of the three systems, i.e., equal importance of digital economy, regional innovation and talent employment, it is taken as $\alpha = \beta = \chi = 1/3$ in this study.

The coupling coordination degree can be evaluated according to the value of coupling coordination degree and the standard is shown in Table 2.

Table 2. Evaluation Standards of Coupling Coordinated Development Type

| Stage | C Range | D Range |
|---|---|---|
| Unbalanced | 0＜C ≤ 0.2 | 0＜D ≤ 0.2 |
| Slightly Unbalanced | 0.2＜C ≤ 0.5 | 0.2＜D ≤ 0.5 |
| Barely Balanced | 0.5＜C ≤ 0.8 | 0.5＜D ≤ 0.8 |
| Superior Balanced | 0.8＜C ≤ 1 | 0.8＜D ≤ 1 |

# 3 Result Analysis

### 3.1 The system score comparison of the Hangzhou Metropolitan Circle

According to the coupling coordination degree calculation above, the scores of the three subsystems of digital economy, regional innovation and talent employment in the Hangzhou Metropolitan Circle are obtained. See Figure 4 for details:

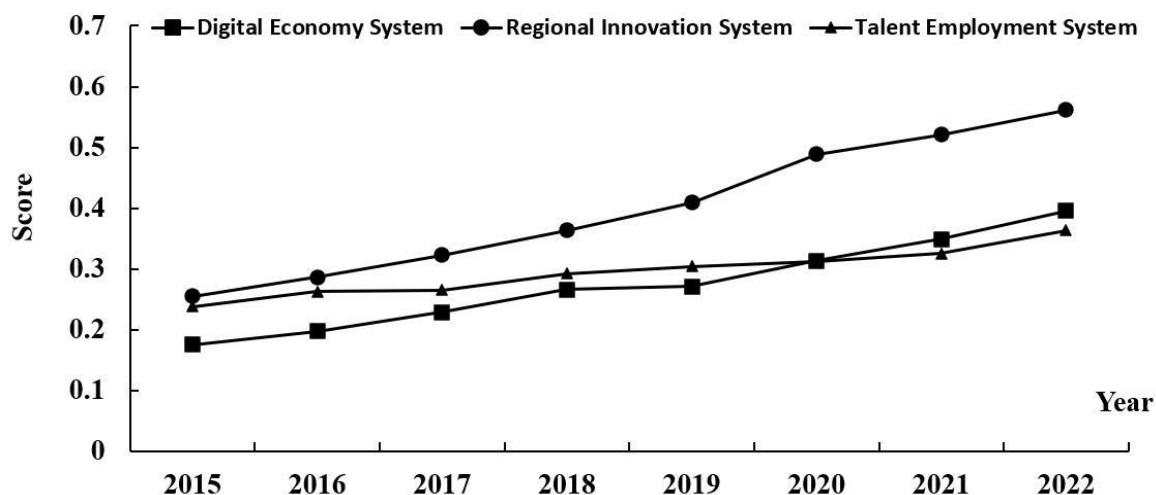

Figure 4. Comprehensive development level in the digital economy, regional innovation and talent employment systems of the Hangzhou Metropolitan Circle

From Figure 3, we find that the regional innovation system is the best among the three systems. From 2015 to 2022, the regional innovation system has always maintained a leading position, leading the comprehensive development of the Hangzhou Metropolitan Circle system. Among them, the regional innovation system began to grow rapidly in 2019, from 0.408 in 2019 to 0.561 in 2022.At the same time, the digital economy system has also begun to grow by a large margin, which shows that regional innovation and digital economy are closely related to each other. The development of talent employment system is the gentlest among the three subsystems. Since 2020, this system has been surpassed by the digital economy system and become the most backward in the three systems. In the perspective of individual city, the performance of the four cities in each subsystem is different. From the diagram of the three subsystems below, we can see that Hangzhou is significantly better than the other three cities.

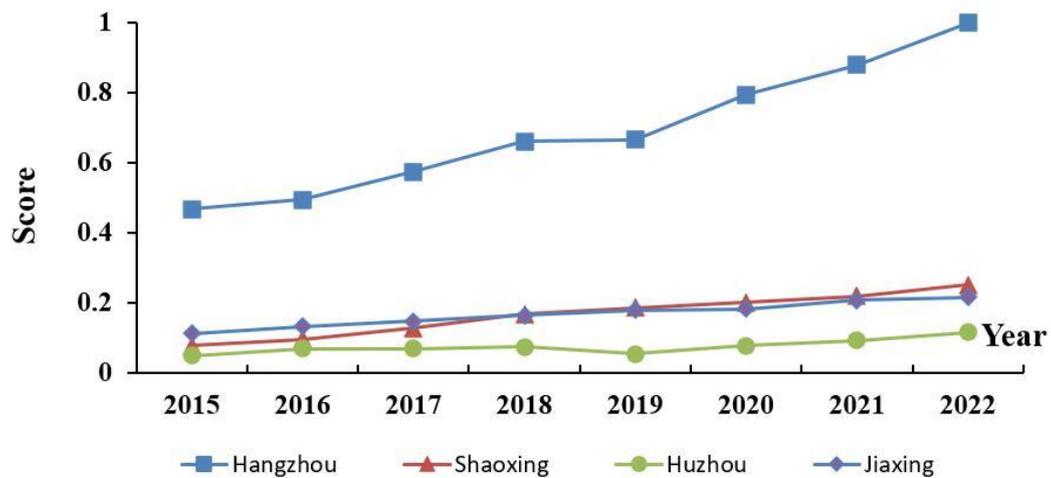

Figure 5. The digital economy system

(1) The digital economy system shows that Hangzhou has significant advantages, ahead of the other three cities. Hangzhou started to develop in 2019, and by 2022, it had reached very close to perfect coordination where D equals 0.99, far ahead of the other three cities. It can be seen that the digital economy has developed into a comparative advantage industry in Hangzhou. A batch of Internet crocodiles represented by Alibaba have gradually become the representatives of new Hangzhou merchants. E-commerce, cultural and creative, financial services, software, the digital economy industries are gradually becoming the leading industries of innovation and development in the urban network of Hangzhou. Jiaxing and Shaoxing have relatively low level of digital economy gathering, but they are growing steadily.

According to the digital economy system, Hangzhou has major advantages over the other three cities. Hangzhou began to develop in 2019, and by 2022, it had reached a level of coordination that was extremely close to ideal, with D equaling 0.99, considerably ahead of the other three cities. It is clear that the digital economy has evolved into a competitive advantage business in Hangzhou. Alibaba's Internet crocodiles have progressively become the representatives of new

Hangzhou merchants. E-commerce, cultural and creative industries, financial services, software, and the digital economy are gradually becoming the key industries of innovation and development in Hangzhou's urban network. Jiaxing and Shaoxing have a low degree of digital economy collecting, but they are steadily developing.

Huzhou has the lowest level of digital economy gathering in the four cities. From the analysis, the four cities in the Hangzhou Metropolitan Circle have different element endowments of the digital economy, which leads to an unbalanced development level in the digital economy bureaus in the region. Among them, digital economy infrastructure construction, foreign investment and other factors have a greater impact on digital economy, but digital economy infrastructure construction may be hardware-biased and ignore the development of supporting software, resulting in the role of digital economy facilities not being fully developed.

In the four cities, Huzhou has the lowest level of digital economy gathering. According to the analysis, the four cities in the Hangzhou Metropolitan Circle have diverse digital economy element endowments, resulting in an unequal development level in the region's digital economy bureaus. Digital economy infrastructure construction, foreign investment, and other factors have a greater impact on the digital economy; however, digital economy infrastructure construction may be hardware-biased and ignore the development of supporting software, resulting in the role of digital economy facilities not being fully developed.

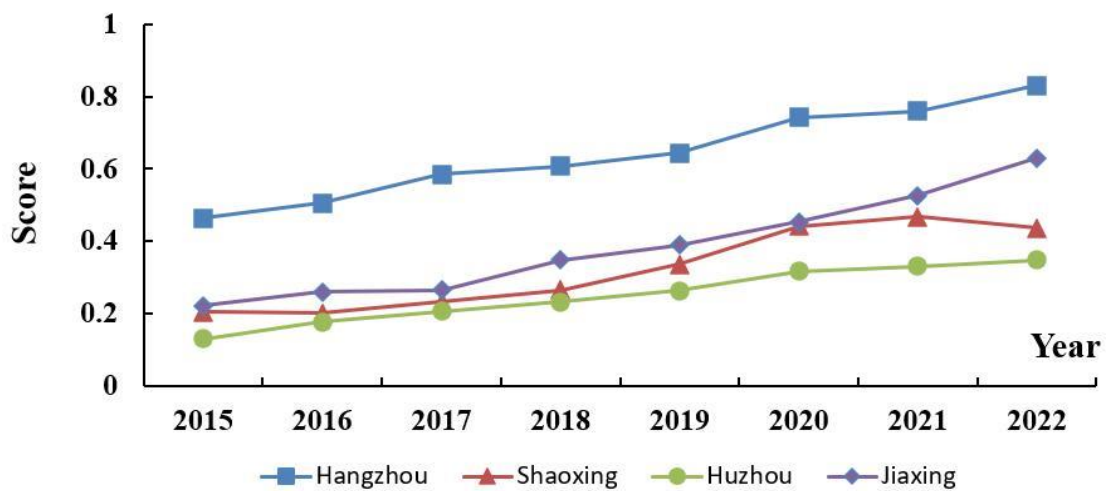

Figure 6. The digital economy system

(2) The regional innovation system has the largest overall increase and the smallest regional differences among the three subsystems. According to the performance of the four cities, Hangzhou is still ahead of the other three cities. Since 2020 especially, Hangzhou has accelerated the "two zones" operation of the cross-border e-commerce Comprehensive Experimental Zone and the national autonomous innovation demonstration zone. The overlapping effect of the free trade policy and innovation policy has been brought into play, which makes the regional innovation advantage of Hangzhou more striking. Among the three subsystems, the regional innovation system has the greatest overall rise and the smallest regional differences. Hangzhou is still ahead of the other three cities in terms of performance. Hangzhou has increased the "two zones" operation of the cross-border e-commerce Comprehensive Experimental Zone and the national autonomous innovation demonstration zone, especially since 2020. The overlapping effect of the free trade policy and the innovation policy has been brought into play, emphasizing Hangzhou's regional innovation advantage.

In addition, a batch of innovative digital economy spaces, such as Dream Town, Yunqi Town, and Zhejiang Science and Technology Park, which integrates industrial functions, cultural functions and community functions, have gradually become an important platform for the gathering of projects, talents and capital, and have provided vitality for regional innovation in Hangzhou Metropolitan Circle. Unlike the digital economy system, several other cities in the system perform better. Among them, Jiaxing's development momentum is only second to Hangzhou's, reaching 0.630 by 2022, and the gap with Hangzhou is narrowing gradually. Shaoxing experienced a rapid increase from 2018 to 2020, followed by a drop in innovation system scores.

Furthermore, a number of innovative digital economy spaces, such as Dream Town, Yunqi Town, and Zhejiang Science and Technology Park, which integrate industrial, cultural, and community functions, have gradually become an important platform for the gathering of projects, talents, and capital, and have provided vitality for regional innovation in the Hangzhou Metropolitan Circle. Several other cities in the system outperform the digital economy system. Jiaxing's development momentum is only second to Hangzhou's, at 0.630 by 2022, and the gap with Hangzhou is gradually closing. Shaoxing witnessed a significant gain from 2018 to 2020, followed by a decline in innovation system scores.

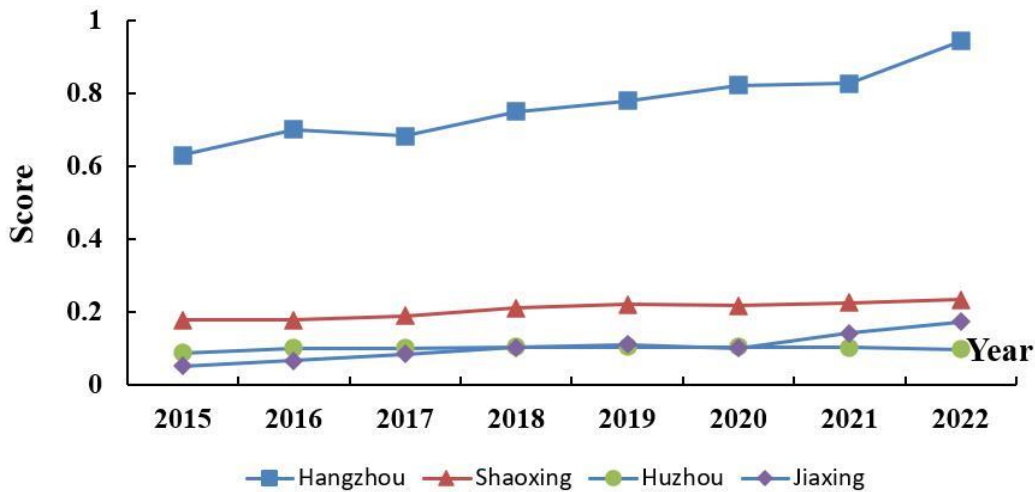
Figure 7. The digital economy system

(3) The development of talent employment system is relatively slow and this system was basically in a stagnant state from 2015 to 2019.As far as the cities are concerned, the regional differences of this system are most obvious. In 2022, Hangzhou's talent employment index reached 0.944, which is far ahead of the rest, and Shaoxing, Jiaxing and Huzhou hovered around 0.2, which greatly reduced the development level of the whole system. In 2022, 218 thousand Hangzhou digital economy workers accounted for 76.9% of the Hangzhou Metropolitan Circle. It can be seen that nearly 80% of the digital economy workers were concentrated in Hangzhou.

The development of the talent employment system has been rather gradual, and the system has been essentially stationary from 2015 to 2019.The regional variances of this system are particularly visible in the cities. In 2022, Hangzhou's talent employment index reached 0.944, substantially outpacing the rest, while Shaoxing, Jiaxing, and Huzhou stayed at 0.2, significantly lowering the whole system's development level. In 2022, 218 thousand Hangzhou digital economy workers accounted for 76.9% of the Hangzhou Metropolitan Circle. It can be seen that roughly 80% of digital economy personnel were centered in Hangzhou.

Among them, the two international scientific and technological innovation bases of the Hangzhou High-tech Zone and Future Science and Technology City have the greatest attraction for high-end talent in the digital economy industry such as "Ali Department", "Zhejiang Business Department", "University Department", and "Return System". Although Shaoxing is close to Hangzhou, the number of employees in the digital economy industry there is lower than the average level in the metropolitan circle, which will inevitably become a problem for Shaoxing in the development of digital economy integration. In addition, the talent in the e-commerce industry is limited by geographic mobility, basically concentrated in Hangzhou, which is also the reason for the slow development of digital economy-related industries in Jiaxing, Huzhou, Shaoxing and other regions.

Among them, the two international scientific and technological innovation bases of the Hangzhou High-tech Zone and Future Science and Technology City have the greatest attraction for high-end talent in the digital economy industry, such as "Ali Department", "Zhejiang Business Department", "University Department", and "Return System". Although Shaoxing is adjacent to Hangzhou, the number of employees in the digital economy industry there is lower than the average level in the metropolitan circle, which will surely become an issue for Shaoxing in the growth of digital economy integration. Furthermore, talent in the e-commerce industry is geographically constrained, primarily concentrated in Hangzhou, which is also the cause for the delayed development of digital economy-related industries in Jiaxing, Huzhou, Shaoxing, and other places.

## 3.2 Measurement of the Coupled and Coordinated Development Level of the Hangzhou Metropolitan Circle

Based on the spatial coupling coordination degree model, this paper calculates the spatial coupling coordination degree of digital economy - regional innovation - talent employment system in the four cities of Hangzhou, Shaoxing, Jiaxing and Huzhou from 2015 to 2022. Table 3 lists the degree of spatial coupling and the degree of spatial coupling coordination.

Table 3. The degree of coupling coordination in the Hangzhou Metropolitan Circle

|  |  | 2015 | 2016 | 2017 | 2018 | 2019 | 2020 | 2021 | 2022 |
|---|---|---|---|---|---|---|---|---|---|
| Hangzhou Metropolitan Circle | C | 0.329 | 0.329 | 0.330 | 0.331 | 0.328 | 0.326 | 0.326 | 0.327 |
|  | D | 0.271 | 0.286 | 0.300 | 0.319 | 0.328 | 0.348 | 0.360 | 0.379 |
| Hangzhou | C | 0.330 | 0.329 | 0.332 | 0.332 | 0.332 | 0.333 | 0.333 | 0.332 |
|  | D | 0.414 | 0.432 | 0.452 | 0.473 | 0.481 | 0.512 | 0.523 | 0.554 |
| Shaoxing | C | 0.306 | 0.317 | 0.324 | 0.328 | 0.323 | 0.312 | 0.312 | 0.312 |
|  | D | 0.217 | 0.224 | 0.243 | 0.265 | 0.283 | 0.299 | 0.308 | 0.314 |
| Jiaxing | C | 0.284 | 0.288 | 0.300 | 0.294 | 0.291 | 0.276 | 0.285 | 0.282 |
|  | D | 0.191 | 0.209 | 0.223 | 0.245 | 0.257 | 0.260 | 0.288 | 0.309 |
|  | C | 0.308 | 0.309 | 0.300 | 0.295 | 0.270 | 0.275 | 0.279 | 0.282 |

| Huzhou | D | 0.166 | 0.190 | 0.194 | 0.201 | 0.195 | 0.214 | 0.221 | 0.230 |

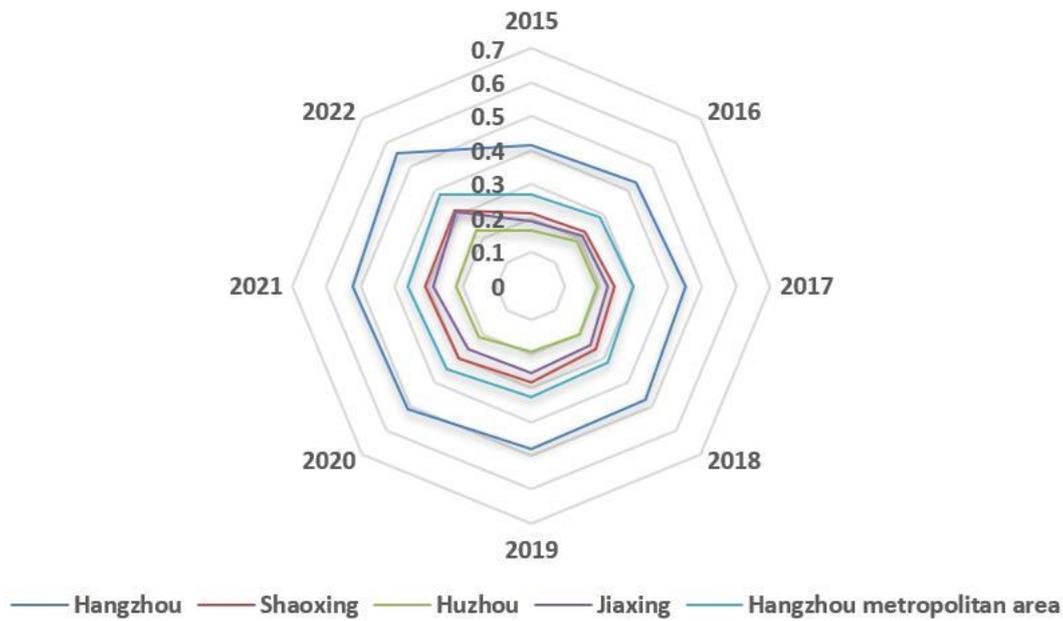

Figure 8. The coupling coordination degree of digital economy - regional innovation - talent employment system of the Hangzhou Metropolitan Circle

Through the construction of the coupling degree function and the coupling coordination model between the three systems of digital economy, we find that from 2015 to 2022, the level of digital economy, regional innovation and talent employment system in the Hangzhou Metropolitan Circle showed a steady upward trend, but the level of coupled coordinated development was relatively flat and only occurs from the non-coordinated period to the barely coordinated period. Among them, the coupling coordination degree is the highest in 2022, which achieves the moderate coordination coupling of digital economy, regional innovation and talent employment. However, for individual cities in the metropolitan circle, only Hangzhou has reached a benign coordination coupling since 2020, while most cities are in a low or moderate coordination coupling stage. We classify the degree of coupling and coordination into three types according to the different level of digital economy, regional innovation and the development of the talent employment system, i.e., the "information lag" type when information aggregation comprehensive development index $f(E)$ is lagging behind the other two systems, the; "innovation lag" type when the comprehensive development index $g(U)$ of regional innovation is the smallest and the "talent lag" type when the comprehensive development index $m(X)$ of employment for talent is the smallest. Based on this, the influencing factors of digital economy, regional innovation and development of talent employment in the Hangzhou Metropolitan Circle from 2015 to 2022 are analyzed (Table 4).

Table 4. Coupling coordination type of digital economy, regional innovation and talent employment in the Hangzhou Metropolitan Circle 2015 - 2022

| Year | Type | Year | Type |
| --- | --- | --- | --- |
| 2015 | Information-Lag | 2019 | Information-Lag |
| 2016 | Information-Lag | 2020 | Talent-lag |
| 2017 | Information-Lag | 2021 | Talent-lag |
| 2018 | Information-Lag | 2022 | Talent-lag |

(1) Urban development in the Hangzhou Metropolitan Circle is not balanced, and the spatial pattern of the degree of coupling and coordination shows a large regional deviation. From 2015 to 2021, the coupling coordination degree of Hangzhou city had been steadily increasing, with a sharp increase in 2022, as it reached the Superior Balanced stage. However, Shaoxing and Jiaxing were in the low-level coupling stage before 2020 and did not reach moderate coordination coupling until around 2021. In recent years, Hangzhou has been vigorously cultivating innovative subjects and gathering innovative elements by building an "Internet+" innovation and entrepreneurship center with global influence. Hangzhou have 85 national level maker spaces and 57 national level incubators by 2022. More than 17,000 technological enterprises have established themselves in Hangzhou, directly employing over 150,000 people. The number of national-level incubators ranked first in provincial capital cities and sub-provincial cities, which shows that the series of support and reward measures issued by Hangzhou had an important role in promoting the cultivation of innovation subjects, platform construction, personnel training, and base construction. For example, Hangzhou set up a small and micro-enterprises innovation base city demonstration "service voucher" and "activity voucher" in 2021. These special funds are used to support small and micro-enterprises and college students in the fields of digital economy, literary creation, science and technology, e-commerce, and so on, and the results of this innovation are remarkable. However, Hangzhou's radiation effect on surrounding cities is not significant.

(2) The Hangzhou Metropolitan Circle has a low degree of coordinated development as a result of "information lag" and "talent lag". The regional innovation system in the three subsystems has always maintained a leading position and leading the comprehensive development of the Hangzhou metropolitan circle system, while the digital economy and talent employment system shows lagging development at different stages. Before 2019, the level of digital economy system development was low, and the system presented an "information lag" type. But after 2019, the digital economy began to develop, the digital economy system steadily improved, and the system presented a "talent lag" type. Dong &Yang (2018) believe that competition also exists in the development of the metropolitan circle, with every metropolitan circle in the Yangtze River Delta "robbing talents" and "robbing resources", and the competition is fierce. The Hangzhou Metropolitan Circle, to whose north is Shanghai, an international metropolis, has a strong attraction for talent, information, funds, and other elements. If the power of Hangzhou Metropolitan Circle is insufficient, the spatial pattern of industry will be unstable and the gathering effect not strong, and the talent and capital elements will easily flow to Shanghai. Similarly, the Ningbo metropolitan circle in the southeast has a strong strength in port logistics, foreign trade and smart manufacturing, and its influence is impressive as well.

(3) The radiation effect of high-quality resources cannot be shown as the resource sharing of Hangzhou Metropolitan Circle is not strong. From 2015 to 2022, Huzhou was in the stage of low coordination coupling, which indicates that the three subsystems of the Huzhou regional digital economy - regional innovation - talent employment system are not coordinated and unstable. The openness of the regional economy makes the digital economy emphasize industry collaboration, value chain interaction, and resource sharing strongly (Scandura, 2016). Compared with the other three cities, Huzhou has a relatively small number of colleges and universities and graduates, and the instability of the talent system is more pronounced. To a certain extent, this also indicates that the high-quality resources in the Hangzhou Metropolitan Circle have not spread well to Huzhou. Huzhou still face the difficulties of gathering high-end talent and other elements, and the problems of high added value but slow development of the high-tech industries. As a result, how to promote the spread of the superior resources of the central city to the surrounding circles, and how to maximize the "1+3" greater than "4" clustering effect is an urgent problem to be solved.

(4) The Hangzhou Metropolitan Circle has formed a number of characteristic industrial space, and an industrial synergistic ecosphere has gradually formed. Characteristically, towns play an important role in supporting the formation of the spatial pattern of "big city town" in the metropolitan network by creating a good industrial ecosphere, attracting new formats, new modes and new talent gathering, and promoting the development of characteristic industries and regional innovation system through collaborative innovation. These feature towns connect with the core circle of the city through efficient information networks and the developed transportation network, forming an industrial collaborative innovation system around the core circle of the city, creating a broad dream-by-dream stage for young people, and providing a good business environment for entrepreneurs (Yang,2016).

## 3.3 Suggestions on the Coupled and Coordinated Development of the Hangzhou Metropolitan Circle

Although the innovation-oriented growth mode has contributed to the Hangzhou Metropolitan Circle's "digital economy-regional innovation-talent employment" system, there is still some unsuitable and uneven development among the three subsystems. As a result, the Hangzhou Metropolitan Circle's system development must take appropriate improvement measures in order to continually optimize its industrial spatial layout and achieve better integration in industry, policy, and human resources.

(1) One approach would be to build an industrial collaboration network in the metropolitan circle and promote coordinated industrial development. Building a batch of labs, research institutes, and industrial service platforms with regional cooperation and sharing through the mode of co-construction, sharing, co-management, and sharing between Hangzhou and Shaoxing, Jiaxing, and Huzhou industrial enterprises in high-end elements and high-end value chains, strengthening the construction of an industry collaboration platform in the Metropolitan Circle, and building a batch of labs, research institutes, and industrial service platforms with regional cooperation and sharing. Since 2018, the network's major cities have upheld the concept of "mutual benefit and win-win, common development," strengthened the sharing of scientific and technological resources, and made breakthroughs in high-end feature clustering, industrial upgrading, and regional cooperation.

(2) In the future, they must prioritize the development of a new regional innovation space and industrial growth platform, as well as the construction of a high-level, multi-level, three-dimensional invention gathering place. This will maximize the inventive benefits of the Hangzhou Independent Innovation Demonstration Zone and the Cross-border E-commerce Comprehensive Test Zone, as well as support the development of the "Internet+" innovation and entrepreneurship platform. The network must cultivate and build a batch of high-level and new-form industrial agglomeration spaces with influence and radiation power, using a double-creation demonstration base, high-tech industrial park, higher education park, small parks, mass-creation space, feature town, and future communities as the breaking points. Taking feature town building as an opportunity, it should actively establish an open, diversified, and harmonious town culture, cultivate feature towns into ideal employment, social spaces, and entrepreneurship spaces for young people, and provide a new growth engine for the regional economy.

(3) The Hangzhou Metropolitan Circle should promote regional talent exchange and the construction of a talent development framework. Governments, universities, and businesses must work together to promote the development of high-level research institutions, such as West Lake University and the Alibaba e-commerce high-tech research and development center, and to accelerate the development of frontier disciplines and fields such as digital economy and information technology. Additional measures should focus on increasing local talent training in information specialties such as computers, communications, software, and e-commerce, establishing a mechanism for mutual selection and recognition of academic credits among college students in metropolitan circles, encouraging the sharing of high-quality vocational education resources across all Hangzhou Metropolitan Circles, and strengthening talent exchange.

## 4. Conclusion

The Hangzhou Metropolitan Circle has a high level of economic development. In order to study the interactive relationship among digital economy, regional innovation and talent development systems, this study builds a model of the coupling degree of coordination, and uses multidimensional indicators to carry out an empirical study on digital economy, regional innovation and the coordinated development of talent employment in the Hangzhou Metropolitan Circle from 2015 to 2022. The results show that the development of the Hangzhou Metropolitan Circle has gone through the stage of unbalanced to balanced, and its development is relatively smooth. Only Hangzhou has a superior coordinated development and good coupling and coordination among the four major cities, while the other three cities have different degrees of "information lag" and "talent lag", which shows that the development of information, talent and innovation resources in the Hangzhou Metropolitan Circle is still unbalanced and inconsistent. This unbalance and inconsistency can easily weaken the influence and correlation between cities in the metropolitan circle and needs to be improved and adjusted in many ways.

In the perspective of policy implication, the Hangzhou Metropolitan Circle needs to establish the industrial advantages and characteristics of the closely related and linked counties (cities and districts). It should take the digital economy as the leading factor, speed up the development of key projects of digital economy, increase the training of talent in the field of digital economy industry. This will promote the development of industries in the Hangzhou Metropolitan Circle towards informatization, technology and high-end. This study covered four cites in the Hangzhou Metropolitan Circle in period of 2015-2022 and the data sample is limited. In the further research, this study would expand the sample by adding other neighboring metropolitan circle for a comparison purpose.